\documentclass[letterpaper, 10 pt, conference]{ieeeconf} 
\IEEEoverridecommandlockouts                     
\usepackage{amsmath} 
\usepackage{amssymb}  
\usepackage{graphicx}
\usepackage{epstopdf}
\usepackage{amsmath}
\usepackage{mathtools}
\usepackage{dsfont}
\let\labelindent\relax
\usepackage{enumitem}
\usepackage{tikz}
\usepackage{bm}
\usepackage{siunitx}
\usepackage{hyperref}
\DeclareMathOperator*{\argmin}{arg\,min}

\usepackage{balance}  
\usepackage{algorithm}
\usepackage{setspace}
\usepackage{caption}
\usepackage{subcaption}
\usepackage{varwidth}
\usepackage{cite}
\usepackage[noend]{algpseudocode}

\usepackage[noend]{algpseudocode}
\newtheorem{assumption}{Assumption}
\newtheorem{remark}{Remark}
\newtheorem{theorem}{Theorem}

\title{\LARGE \bf
Data-Driven Hierarchical Predictive Learning in Unknown Environments}

\author{Charlott Vallon and Francesco Borrelli
\thanks{*This research was sustained in part by fellowship support from the National Physical Science Consortium, the National Institute of Standards and Technology, and by the grant NSF-1931853. }
\thanks{$^{1}$Charlott Vallon, and Francesco Borrelli are with the Department of Mechanical Engineering, University of California, Berkeley, Berkeley, CA 94701, USA
        {\tt\small \{charlottvallon, fborrelli\} $@$ berkeley.edu}}%
}%
\begin{document}

\maketitle
\thispagestyle{empty}
\pagestyle{empty}

\begin{abstract}
We propose a hierarchical learning architecture for predictive control in unknown environments. 
We consider a constrained nonlinear dynamical system and assume the availability of state-input trajectories solving control tasks in different environments. 
A parameterized environment model generates state constraints specific to each task, which are satisfied by the stored trajectories.
Our goal is to find a feasible trajectory for a new task in an unknown environment. 
From stored data, we learn strategies in the form of target sets in a reduced-order state space.
These strategies are applied to the new task in real-time using a local forecast of the new environment, and the resulting output is used as a terminal region by a low-level receding horizon controller.
We show how to \emph{i)} design the target sets from past data and then \emph{ii)} incorporate them into a model predictive control scheme with shifting horizon that ensures safety of the closed-loop system when performing the new task. 
We prove the feasibility of the resulting control policy, and verify the proposed method in a robotic path planning application.
\end{abstract}


\section{Introduction}

Classical Iterative Learning Controllers (ILCs) aim to improve a system's closed-loop reference tracking performance at each iteration of a repeated task \cite{AA, batchILC, newcite}.
Recent work has also explored reference-free ILC for applications whose goals are better defined through a performance metric, such as racing autonomously around a track or harvesting wind energy \cite{rosolia2017autonomousrace, melanie, vermillion}.

In general, the learned ILC policy is not cost-effective, or even feasible, if the task environment changes \cite{rosolia2016learning, vallon2019task}. 
In the Artificial Intelligence and Reinforcement Learning communities, the ability to generate a control policy which performs well under different environment conditions is a common challenge, often referred to as generalization or transfer learning \cite{transferlearning, weiss2016survey}.
These data-driven approaches typically focus on minimizing the performance loss between solving tasks in the original and new environment, rather than guaranteeing feasibility of the policy in a new environment.

Methods that do guarantee feasibility are generally model-based \cite{bertsimas2018voice, hewing2019cautious, stolle, berenson, pereida2018data}. Approaches have been proposed for autonomous vehicles \cite{zhi2019octnet} and robotic manipulation applications \cite{probabilisticprimitives, fitzgerald, vallon2019task}.
These strategies often require maintaining a trajectory library, and adapting the stored trajectories online to the new constraints of the changed tasks, which can be both time-consuming and computationally expensive.

This article proposes a data-driven method for tackling a simple abstraction of the ``changing environment" control problem.
We consider availability of state-input trajectories which solve a set of control tasks $\{\mathcal{T}^1$,\ldots,$\mathcal{T}^n\}$.
A successful execution of the $i^\mathrm{th}$ control task $\mathcal{T}^i$ is defined as a trajectory of states and inputs evolving according to a nonlinear difference equation, satisfying system state and input constraints, and satisfying constraints imposed by the task environment.
In  each of the $n$ control tasks the system model, system constraints and objective function are identical. However, a parameterized  environment descriptor function $\bm{\theta}$ generates a set of state constraints specific to each task $\mathcal{T}^i$, where
$\bm{\theta}=\bm{\theta}^i$. 

Our goal is to use stored executions of $n$ previous control tasks in order to complete a successful execution of a new task
$\mathcal{T}^{n+1}$ in a new environment described by $\bm{\theta}^{n+1}$.
This paper presents a hierarchical predictive learning architecture to solve such a problem. Specifically, we use stored data to design strategy sets in reduced-dimension state space. These strategy sets represent high-level strategies learned from previous control tasks, and are used as waypoint regions in a receding horizon control. In this paper, we:
\begin{enumerate}
    \item propose interpreting ``strategies" as sets in reduced-dimension state space, referred to as ``strategy sets", 
    \item show how to design strategy sets from past data,
    \item demonstrate how to incorporate the strategy sets into a receding horizon control, 
    \item prove the proposed method will lead to a successful execution of the new control task, and 
    \item apply the controller to a robotic path planning example.
\end{enumerate}

\section{Problem Formulation}

We consider a discrete-time system with dynamical model
\begin{align}\label{eq:VehicleModel}
    x_{k+1} &= f(x_k, u_k),
\end{align}
subject to system state and input constraints
\begin{align}\label{eq:VehicleModelConstr}
    x_k \in \mathcal{X},~ u_k \in \mathcal{U}.
\end{align}
The vectors $x_k \in \mathbb{R}^{n_x}$ and $u_k \in \mathbb{R}^{n_u}$ collect the states and inputs at time $k$.

\subsection{Task Environments}
\label{ssec:tasksandexecutions}
The system (\ref{eq:VehicleModel}) solves a series of $n$ control tasks $\{\mathcal{T}^1, \dots, \mathcal{T}^n\}$ in different task environments parameterized by $\{{\bm{\theta}}^1, \dots, {\bm{\theta}}^n\}$. For each task $\mathcal{T}^i$, the environment descriptor function ${\bm{\theta}}^i$ maps the state $x_k$ at time $k$ to a description of the local task environment
$\bm{\theta}^i(x_k, k)$.
Examples of ${\bm{\theta}^i(x_k, k)}$ include camera images, the coefficients of a polynomial describing a race track lane boundaries, or simple waypoints for tracking.

In each control task the system model (\ref{eq:VehicleModel}) and constraints (\ref{eq:VehicleModelConstr}) are identical. However, the environment descriptor function generates additional local environment-specific state constraints, denoted $E({\bm{\theta}}^i(x_k, k))$. In task $\mathcal{T}^i$, we write the combined system and environment constraints as 
\begin{align}\label{eq:taskconstraints}
    x_k \in \mathcal{X}({\bm{\theta}}^i, x_k, k),
\end{align}
where
\begin{align}
    \mathcal{X}({\bm{\theta}}^i, x_k, k) =  E({\bm{\theta}}^i(x_k, k)) \cap \mathcal{X}. \nonumber
\end{align}
For notational simplicity, we refer to the combined system and environment constraints (\ref{eq:taskconstraints}) as $\mathcal{X}({\bm{\theta}}^i)$.

\subsection{Task Executions}
A feasible execution of the task $\mathcal{T}^i$ in environment $E({\bm{\theta}}^i)$ is defined as a pair of state and input trajectories
    \begin{align}\label{eq:iterationvecs}
\mathrm{Ex}(\mathcal{T}^i, {\bm{\theta}}^i) & = [{\bf{x}}^i, {\bf{u}}^i] \\ 
    {\bf{x}}^i & = [x^i_{0},x^i_{1},...,x^i_{D^i}], ~x^i_k \in \mathcal{X}({\bm{\theta}}^i)~~ \forall k \in [0, {D^i}], \nonumber\\ & ~~~~~~~~~~~~~~~~~~~~~~~ x_{D^i} \in \mathcal{P}({\bm{\theta}}^i),\nonumber\\
    {\bf{u}}^i & = [u^i_0,u^i_1,...,u^i_{D^i}], ~u_k \in \mathcal{U}~~~~~~~ \forall k \in [0, D^i],\nonumber
    \end{align}
where ${\bf{u}}^i$ collects the inputs applied to the system (\ref{eq:VehicleModel}) and ${\bf{x}}^i$ is the resulting state evolution. $D^i$ is the duration of the execution of task $\mathcal{T}^i$. The final state of a feasible task execution, $x_{D^i}$, is in the task target set $\mathcal{P}({\bm{\theta_i}}) \subset \mathcal{X}({\bm{\theta}}^i)$.

\textit{Problem Definition:}
Given a dynamical model (\ref{eq:VehicleModel}) with state and input constraints (\ref{eq:VehicleModelConstr}),~(\ref{eq:taskconstraints}), and a collection of feasible executions (\ref{eq:iterationvecs}) that solve a series of $n$ control tasks, $\{\mathrm{Ex}(\mathcal{T}^1, {\bm{\theta}}^1), ..., \mathrm{Ex}(\mathcal{T}^n, {\bm{\theta}}^n) \}$, our aim is to find an execution for a new task in an unseen environment: $\mathrm{Ex}(\mathcal{T}^{n+1}, \bm{\theta}^{n+1})$.

\begin{remark}
For notational simplicity, we write that the collected data set contains one execution from each task.
However, in practice one may collect multiple executions of each task $\mathcal{T}^i$. In this case, the executions can simply be stacked, and the procedure proposed in this paper can proceed as described.
\end{remark}

\subsection{Environment Forecasts}

The control architecture proposed in this paper relies on forecasts of the task environment.
At time $k$, we use forecasts of the system state across a horizon $N$, $\hat{x}^i_{k:k+N}$, to predict the corresponding environments $[{\bm{\theta}}^i(\hat{x}_k, k), \dots, {\bm{\theta}}^i(\hat{x}_{k+N}, k+N)]$. 
For notational simplicity, we write this environment forecast as ${{\theta}}^i_{k : k+N}$.

\section{Hierarchical Predictive Learning Control}\label{sec:hpl}
We propose a data-driven controller that uses stored executions from previous tasks, $\{\mathrm{Ex}(\mathcal{T}^1, \bm{\theta}^1), \dots,\mathrm{Ex}(\mathcal{T}^n, \bm{\theta}^n)\}$, to find an execution for a new task in a new environment, $\mathrm{Ex}(\mathcal{T}^{n+1},\bm{\theta}^{n+1})$.
Instead of simply adapting the stored executions from previous tasks to the constraints of the new task, our aim is to learn a generalizable and interpretable \textit{strategy} from past tasks, and apply it to the new task. 

\subsection{Approach}\label{ssec:approach}
Our approach is inspired by how navigation tasks are typically explained to humans, who can easily generalize learning to new environments. For example, if a human has learned to race a vehicle by driving around a single track, they can easily adapt their learned strategy when racing a new track. 
Based on real-life intuition, we propose three principles of strategy:
\begin{enumerate}
    \item Strategies are a function of a local environment forecast.
    \item Strategies work in a reduced-order state space.
    \item Strategies provide target regions in the (reduced-order) state space for which to aim. 
\end{enumerate}
The control architecture proposed in this paper formalizes the above principles of strategy, and shows how to incorporate this strategy framework into a hierarchical learning control.
In particular, we focus on two aspects. First, we show how to learn generalizable strategies from stored executions of previous control tasks.
Second, we show how to integrate the learned strategies in an MPC framework so as to guarantee feasibility when solving a new control task in real-time.

\begin{figure}[h!]
    \centering
    \includegraphics[width=0.9\columnwidth]{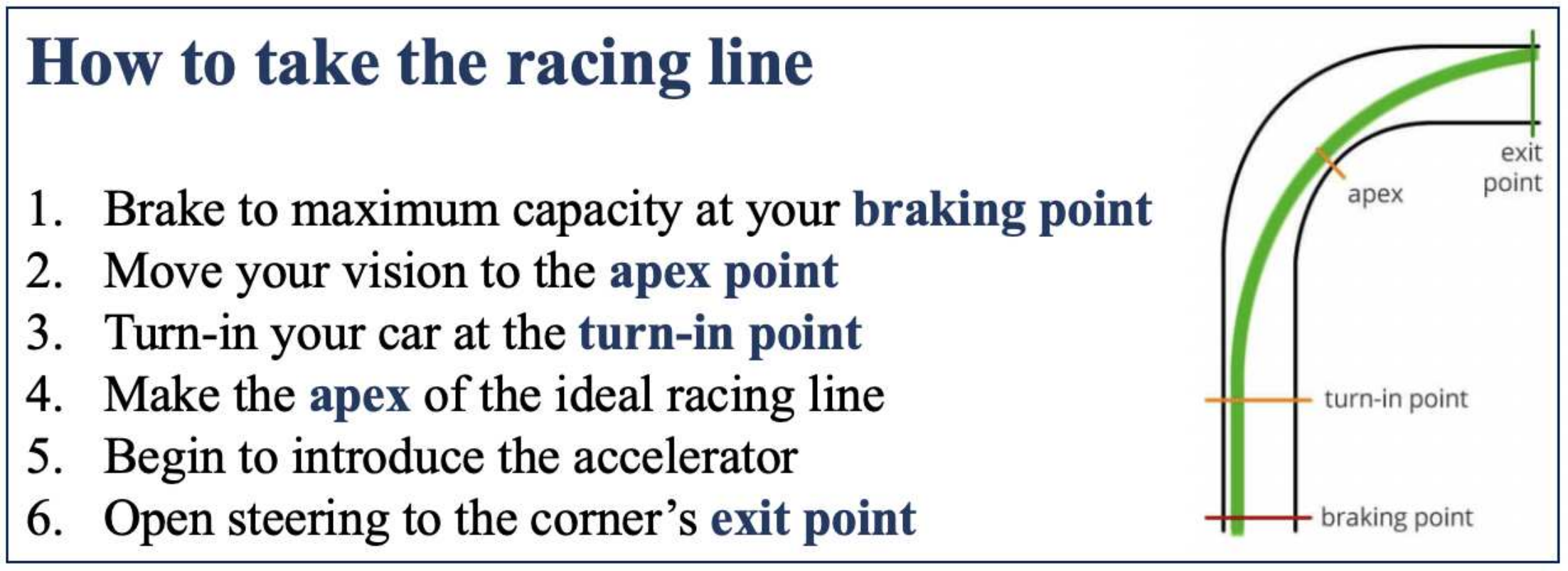}
    \caption{Sample of strategies taught at various online racing schools.}
    \label{fig:racing_strategies}
\end{figure}

\textit{Vehicle Racing Example:} 
Consider, for example, learning how to race a vehicle around a track. A snippet of common racing strategies\footnote{As taught at online racing schools such as Driver 61: \href{https://driver61.com/uni/racing-line/}{\url{https://driver61.com/uni/racing-line/}}}  taught to new racers is depicted in Fig.~\ref{fig:racing_strategies}.
The environmental constraints imposed on the vehicle are parameterized by the race track curvature.
This environmental descriptor gives rise to physical areas in state space with respect to which racing rules are then explained, e.g. ``cut the curve at the apex and aim for the outside of the straightaway." 
The strategies consist of sections of the track towards which to aim the vehicle as well as acceleration profiles to apply along the way.
The strategies are explained using only a subset of the state space: the distance from the centerline. 
Given guidelines on this subset, the driver is free to adjust other states and inputs such as vehicle velocity and steering as necessary.

\subsection{Implementation}
Hierarchical Predictive Learning (HPL) is a data-driven control scheme based on high-level strategies, learned from previous executions of different tasks according to our principles of strategy. The HPL controller modifies its behavior whenever new strategies become applicable, and operates in coordination with a safety controller to ensure constraint satisfaction at all future time steps.

At each time $k$, an $N$-step local environment forecast ${{\theta}}^i_{k:k+N}$ is used to determine if a new high-level control strategy is available. 
A strategy consists of state and input sets in reduced dimensions, $\tilde{\mathcal{X}}_{k+T}$ and $\tilde{\mathcal{U}}_{k:k+T}$, that provide a set towards which to steer the system in the next $T$ timesteps, as well as input guidelines for getting there.
The strategy sets are used to construct a target set in the full state space, ${\mathcal{X}}_{k+T}$.
Lastly, a receding horizon controller calculates a low-level input to get the system to the target set.
In the following sections, we explain each of the three hierarchy levels in detail and prove the feasibility of the resulting control law.

We note the difference between the environment forecast horizon $N$ and the strategy prediction horizon $T$. The two need not be the same, and it is reasonable to select $N>T$.

\section{Learning Strategies From Data}\label{sec:strategyfinder}
This section addresses the first aim of our paper: learning generalizable strategies from previously solved tasks. 
There are several ways of learning strategies, including model-based methods that use an explicit model for how variations in task environments affect the optimal control input \cite{bertsimas2018voice}. 
In this work we instead opt for a data-driven approach, using stored executions (\ref{eq:iterationvecs}) that solved related tasks. 

We propose using Gaussian Processes (GPs) to learn a mapping from a current system state and environment forecast, $(x_k, \theta^i_{k:k+N})$, to hyperrectangular state and input constraint  sets in reduced dimension, $(\tilde{\mathcal{X}}_{k+T}, \tilde{\mathcal{U}}_{k:k+T})$, known as strategy sets.
The GPs are trained \textit{offline}, and the strategy sets are built \textit{online} during the execution of a new control task using the estimated mean and standard deviation provided by the GPs. 
A review of GPs in control is provided in \cite{kocijan2016modelling}, and implementation details in the Appendix. 

\subsection{Training the GP (Offline)}\label{ssec:trainingGP}

\subsubsection{GPs in MPC}
GPs have been used in recent predictive control literature as data-driven estimates of unknown nonlinear dynamics \cite{hewing2019cautious, learningGaussProcess,klenske2015gaussian}. 
Specifically, GPs are used to approximate vector-valued functions with real (scalar) outputs.
Given training data (input vectors and output values), GPs learn a similarity measure known as ``kernel" between inputs to estimate the true underlying function. 
The kernel represents the learned covariance between two function evaluations. 

\subsubsection{Why use GPs?}
Once a kernel has been learned, the GP can be queried at a new input vector. 
Given an input, the GP returns a one-dimensional Gaussian distribution over output estimates; thus GPs provide a best guess for the output value corresponding to an input \textit{and} a measure of uncertainty about the estimate based on the distribution's standard deviation.
This allows us to gauge how confident the GP is in its prediction at a particular input. 

\textit{Note:} The robust MPC community often prefers stochastic models with bounded support \cite{bujarbaruahAdapFIR} to GPs, in order to have strict safety guarantees. Our approach can be extended to these types of models as well. 

\subsubsection{GPs for Hierarchical Predictive Learning}
We consider strategies to be maps from a state and environment forecast to reduced-dimension strategy sets. The choice of appropriate strategy states and inputs depends on the tasks.
We define the strategy state at each time $k$ as
\begin{align}\label{eq:stateprojection}
    \tilde{x}_k = g(x_k) \in \mathbb{R}^{n_{\tilde{x}}},
\end{align}
where $g(\cdot)$ maps the full-dimensional state $x_k$ into the corresponding lower-dimensional strategy state. Similarly, the strategy inputs are
\begin{align}\label{eq:inputprojection}
    \tilde{u}_k = r(u_k) \in \mathbb{R}^{n_{\tilde{u}}},
\end{align}
where $r(\cdot)$ maps the full-dimensional input at time $k$ into lower-dimensional strategy inputs.
We denote the strategy state and strategy input spaces as
\begin{align}\label{eq:sxset}
    \tilde{\mathcal{X}} &= \{\tilde{x} ~| ~\tilde{x} = g(x),~ x \in \mathcal{X} \},\\ 
    \tilde{\mathcal{U}} &= \{\tilde{u} ~|~ \tilde{u} = r(u),~ u \in \mathcal{U} \}.
\end{align}
In HPL, we train GPs to predict the values of the strategy states (\ref{eq:stateprojection}) and inputs (\ref{eq:inputprojection}) at $T$ timesteps into the future, based on the current state and environment forecast.
Each GP learns to approximate the mapping between a state and environment forecast and a particular strategy state or input:
\begin{align}\label{eq:gpeq}
    (\mu, \sigma^2) = \mathrm{GP}(\tilde{x}_k, \theta^i_{k:k+N}),
\end{align}
where $\mu$ and $\sigma^2$ represent statistics of the Gaussian distribution over strategy state estimates which are used to build strategy sets in Sec.\ref{ssec:stratsets}.
GPs best approximate functions with scalar outputs, so we train one GP for each strategy state and input (a total of $n_{\tilde{x}} + n_{\tilde{u}}$ number of GPs). 

\subsubsection{Building Training Data}
We use stored executions (\ref{eq:iterationvecs}) from previous control tasks to create GP training data. Each GP uses the same training input data. The training output data for each GP contains the strategy state or input the GP is learning to predict. 
After solving $n$ control tasks at least once (see Remark 1), the training data consists of:
\begin{align}\label{eq:dataset}
    \mathcal{D} = \{{\bf{z}} & = [z^1_0, z^1_1, \dots,  z_{D^1-T}^1, z_0^2, \dots, ..., z_{D^n-T}^n] \\ 
    {\bf{y}} & = [y^1_0, y^1_1, \dots, y_{D^n-T}^n]\}. \nonumber
\end{align}
Each input vector $z^i_j$ corresponds to the output $y^i_j$, where
\begin{align}\label{eq:datainput}
    z^i_j = [x^i_j, \theta^i_j, \theta^i_{j+1}, ..., &\theta^i_{j+N}], \\
    & ~ i \in [1,n], ~ j \in [0, D^i - T],\nonumber
\end{align}
and $\theta^i_j$ denotes the local environment at time $k$ of the $i$th control task. For example, this can correspond to the instantaneous track curvature or the camera image recorded at that time step.
The output entry $y^i_j$ contains the value of the strategy state or input of interest at $T$ time steps in the future:
\begin{align}
    y^i_j = \tilde{x}^{i}_{j+T} ~ \mathrm{or} ~ \tilde{u}^{i}_{j+T}&, \nonumber\\
    & ~ i \in [1,n], ~ j \in [0, D^i - T].\nonumber
\end{align}
Again, we emphasize the difference between the environment forecast horizon $N$ and the strategy prediction horizon $T$. 

\subsubsection{Training}
We use the Matlab Statistics and Machine Learning toolbox\footnote{\url{https://www.mathworks.com/help/stats/index.html}} to learn kernel hyperparameters that best match the training data (\ref{eq:dataset}). 
Specifically, we train a squared-exponential kernel with separate length scales for each predictor (see Appendix). New hyperparameters can be learned whenever more executions become available.

\subsection{Constructing the Strategy Sets (Online)} \label{ssec:stratsets}
Once the GPs are trained, they can be used to solve a new task. 
At time $k$ of a new task $\mathcal{T}^{n+1}$, we evaluate the GPs at the new query vector $z^{n+1}_k$, formed as in (\ref{eq:datainput}), to construct hyperrectangular strategy sets in reduced-dimension space.

\begin{figure}
    \centering
    \includegraphics[width = 0.9\columnwidth]{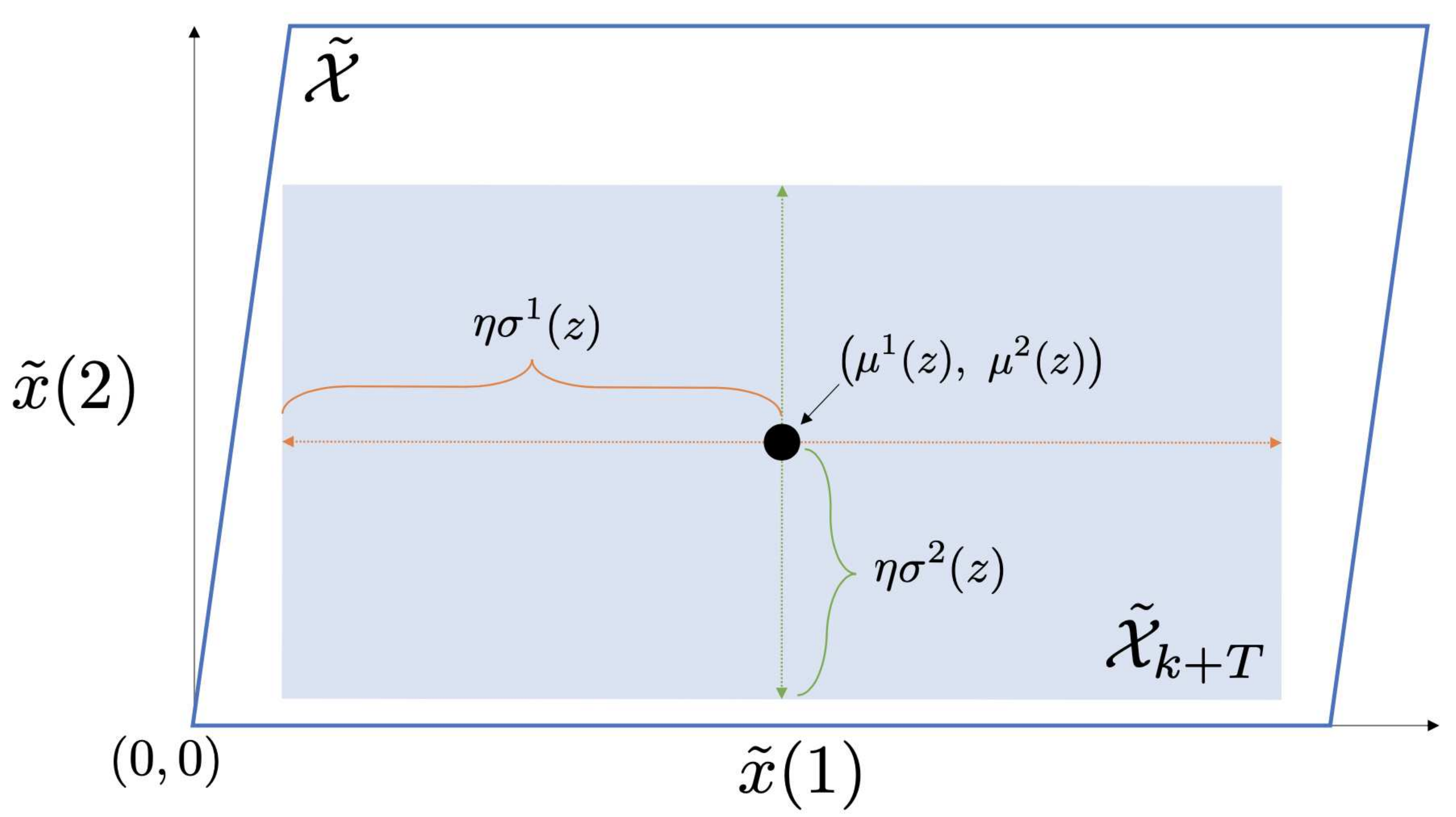}
    \caption{Each dimension $\tilde{x}(i)$ of the strategy set $\tilde{\mathcal{X}}_{k+T}$ is bounded using a GP (\ref{eq:strategystateconstraint}).}
    \label{fig:hyperrect}
\end{figure}

Each GP returns a one-dimensional Gaussian distribution over output scalars, parameterized by a mean $\mu$ and variance $\sigma^2$.
Once means and variances have been determined for each strategy state and input from the learned kernel hyperparameters (see (\ref{eq:meaneval}) in Appendix), we form one-dimensional bounds on each $i$th strategy state and $j$th strategy input as
\begin{align}
    \tilde{\mathcal{X}}_{k+T}(i) &= [\mu^i(z_k^{n+1}) \pm \eta \sigma^i(z_k^{n+1})],~  \forall i \in [1, n_{\tilde{x}}] \label{eq:strategystateconstraint}\\
    \tilde{\mathcal{U}}_{k:k+T}(j) &= [\mu^{j}(z_k^{n+1}) \pm \eta \sigma^{j}(z_k^{n+1})],~  \nonumber \\
    & ~~~~~~~~~~~~~~~~~~~~~~~~~~~\forall j \in [n_{\tilde{x}}+1, n_{\tilde{x}} + n_{\tilde{u}}]. \nonumber
\end{align}
In (\ref{eq:strategystateconstraint}), $\mu^i(\cdot)$ and $\sigma^i(\cdot)$ are the means and standard deviations computed by the $i$th GP. 
The parameter $\eta>0$ determines the size of the range.
When these one-dimensional bounds are combined for all strategy states and strategy inputs, respectively, we form hyperrectangular strategy sets in strategy space, with each dimension constrained according to (\ref{eq:strategystateconstraint}).
An example is shown in Fig.~\ref{fig:hyperrect}.
The hyperrectangular strategy sets are denoted $\tilde{\mathcal{X}}_{k+T}$ and $\tilde{\mathcal{U}}_{k:k+T}$, and indicate where (in strategy space) the system should be in $T$ timesteps and what inputs to apply to get there.

\section{Safely Applying Learned Strategies}\label{sec:strategymanager}

We now address the second aim of our paper: using the strategy sets in a low-level controller while maintaining safety guarantees. 
Our approach consists of \textit{i)} lifting the reduced-dimension strategy sets back into full-dimension, and \textit{ii)} integrating the full-dimension set with a safety controller. The result is a target set that can be used in a low-level MPC controller.

\begin{assumption}\label{ass:safetycontroller}
There exists a safety control policy that can prevent the system (\ref{eq:VehicleModel}) from violating both system- and task-specific environment constraints (\ref{eq:taskconstraints}). 
In particular, there exists a safe set 
\begin{equation}\label{eq:emergencyset}
    \mathcal{X}_E \subseteq \mathcal{X}(\bm{\theta}),
\end{equation}
and a corresponding safety control law
\begin{equation}\label{eq:emergencycontrol}
    u = \pi_{\mathrm{e}}(x, \bm{\theta}),
\end{equation}
such that $\forall x \in \mathcal{X}_E,~ f(x, \pi_{\mathrm{e}}(x, \bm{\theta})) \in \mathcal{X}_E.$
\end{assumption}
\begin{remark}
Given a safety control law (\ref{eq:emergencycontrol}), the safe set (\ref{eq:emergencyset}) may be found using a variety of data-driven methods, including backwards reachability from a known subset of $\mathcal{X}_E$ (such as physical standstill), or sample-based forward reachability from a gridded state space $\mathcal{X}$.
\end{remark}

\subsection{Converting to full-dimensional hyperrectangles} 
At each time $k$ of solving a task $\mathcal{T}^{n+1}$, new reduced-dimension strategy sets (\ref{eq:strategystateconstraint}) are constructed. These strategy sets are converted to a target set in full-dimensional state space, to be used as a terminal constraint in a low-level MPC controller. 
Critically, the target set must belong to the safe set (\ref{eq:emergencyset}). This ensures that once the system has reached the target set, there will always exist at least one feasible input (the safety control (\ref{eq:emergencycontrol})) that allows the system to satisfy all state constraints.
Given strategy sets (\ref{eq:strategystateconstraint}) found at time $k$, we define a corresponding lifted strategy set as:
\begin{align}\label{eq:xt}
    \mathcal{X}_{k+T} = \{x \in \mathcal{X}_E ~| ~g(x) \in \tilde{\mathcal{X}}_{k+T} \},
\end{align}
where $g(x)$ (\ref{eq:stateprojection}) is the projection of the full-dimensional state $x$ onto the set of chosen strategy states. 
$\mathcal{X}_{k+T}$ is a full-dimensional set in which the strategy states (\ref{eq:stateprojection}) are constrained to lie in the strategy sets (\ref{eq:strategystateconstraint}), and the remaining states are constrained such that for any state $x_k \in \mathcal{X}_{k+T}$, the safety control (\ref{eq:emergencycontrol}) can be applied if necessary to ensure constraint satisfaction in future time steps. 

\subsection{Incorporating the Uncertainty Measure}
A benefit of using GPs is that the standard deviation around an estimate may be used to evaluate how confident the GP is in its prediction at a particular input.
At time $k$, consider the uncertainty measure $\mathcal{C}_k$ which is a vector containing the standard deviations of the evaluated GPs:
\begin{align}\label{eq:uncertaintymeasure}
    \mathcal{C}_k = [\sigma^1(z_k^{n+1}), \dots , \sigma^{n_{sx}+n_{su}}(z_k^{n+1})]. 
\end{align}
If the GPs return a strategy set with standard deviations (\ref{eq:uncertaintymeasure}) larger than a chosen threshold $d_{\mathrm{thresh}}$, we may opt not to use this strategy. 
We expect $\mathcal{C}_k > d_{\mathrm{thresh}}$ if:
\begin{enumerate}
    \item the system did not encounter a similar environment forecast in a previous control task, or
    \item in previous control tasks this environment forecast did not lead to a single coherent strategy, resulting in a wide distribution of potential future strategy states.
\end{enumerate}
With high uncertainty measures, the strategy sets are not likely to contain valuable control information for the system. 
In this case, HPL sets the target set to be \textit{empty}: $\mathcal{X}_{k+T} = [~]$.
As explained in Sec.~\ref{sec:lowlevelcontrol}, this results in a horizon shift for the low-level MPC, and the system (\ref{eq:VehicleModel}) re-uses the target set from the previous time step.

\section{Low-level Controller Design}\label{sec:lowlevelcontrol}

The low-level MPC controller is responsible for calculating the input to be applied to the system at each time $k$.
This input is calculated based on the sequence of target sets (\ref{eq:xt}) found during the last $T$ timesteps.

\subsection{Target Set List}
At each time $k$ of solving task $\mathcal{T}^{n+1}$, a new target set (\ref{eq:xt}) is constructed by lifting the strategy sets (\ref{eq:strategystateconstraint}). 
However, if the standard deviations (\ref{eq:uncertaintymeasure}) are too high, or there is no feasible input sequence to reach the lifted strategy set $\mathcal{X}_{k+T}$, the target set for time $k$ will be empty: $\mathcal{X}_{k+T} = [~]$.

The ``target set list" keeps track of which target sets (empty or not) were constructed during the most recent $T$ timesteps:
\begin{align}\label{eq:setlist}
    \mathrm{SetList}_k = [{\mathcal{X}}_{k+1}, {\mathcal{X}}_{k+2}, \dots, {\mathcal{X}}_{k+T}].
\end{align}
At each new time step, the first set gets removed and the target set found at the current time step gets appended to the end. 
In this way, the target set list (\ref{eq:setlist}) always maintains exactly $T$ sets, though some (including the last one) may be empty.
This list is used to guide the objective function and constraints of the MPC controller. 

\subsection{Shifting Horizon MPC Formulation}
We formulate an MPC controller to calculate our input at each time step:
\begin{align}\label{eq:lowlevelmpc}
{\bf{u}}^\star(x_k) &= \argmin_{u_{0|k},...,u_{N_k-1|k}} \sum_{j\in \mathcal{S}_k}^{N_k-1}\mathrm{dist}\big(x_{j|k}, \mathcal{X}_{k+j}\big) \\
& ~~~~~~ \mathrm{s.t.}~~ x_{j+1|k} = f(x_{j|k}, u_{j|k}) \nonumber\\ & ~~~~~~~~~~~~~~~ x_{j|k} \in \mathcal{X}(\bm{\theta}^{n+1}) ~~\forall j \in \{ 0, N_M-1\} \nonumber\\
& ~~~~~~~~~~~~~~~ u_{j|k} \in \mathcal{U}~~~~~~~~~~~ \forall j \in \{ 0, N_M-1\}\nonumber\\
& ~~~~~~~~~~~~~~~ x_{N_k|k} \in \mathcal{X}_{k+N_k}\nonumber \\
& ~~~~~~~~~~~~~~~ x_{0|k} = x_k,\nonumber
\end{align}
where $\mathcal{S}_k$ is the set of indices with non-empty target sets, 
\begin{align}
    \mathcal{S}_k = \{s~ |~ \mathrm{notEmpty}(\mathcal{X}_{k+s-1}) \}. \nonumber
\end{align}
The MPC objective function (\ref{eq:lowlevelmpc}) penalizes the Euclidean distance from each predicted state to the target set corresponding to that prediction time. If a smoother cost is desired, the objective could be augmented to take the input effort into account.

The MPC uses a time-varying shifting horizon $0 < N_k \leq T$ that corresponds to the largest time step into the future for which a non-empty target set results in feasibility of (\ref{eq:lowlevelmpc}):
\begin{align}\label{eq:nk}
    {N_k} = \max s~:~ \{s \in \mathcal{S}_k,~ (\ref{eq:lowlevelmpc}) \text{ is feasible with $N_k = s$} \}. 
\end{align}
This ensures that the MPC controller (\ref{eq:lowlevelmpc}) has a non-empty terminal constraint and the optimization problem is feasible. 
To avoid unnecessary repeated computations, all target sets in the target set list (\ref{eq:setlist}) which lead to infeasibility of (\ref{eq:lowlevelmpc}) when used as the terminal constraint are set as \textit{empty} in (\ref{eq:setlist}).
At time step $k$, we apply the first optimal input to the system:
\begin{align}\label{eq:lowlevelinput}
    u_k = u^\star_{0|k}.
\end{align}

\textit{Note:} Here, the target sets (\ref{eq:xt}) are found at the same frequency as the controller (\ref{eq:lowlevelmpc})-(\ref{eq:lowlevelinput}) is updated, but our framework can easily be adapted to use asynchronous loops.

\subsection{Safety Control}
If no target sets in (\ref{eq:setlist}) can be used as a terminal constraint in (\ref{eq:lowlevelmpc}) such that (\ref{eq:lowlevelmpc}) is feasible, all sets in the target set list (\ref{eq:setlist}) will be empty, and the MPC horizon is $N_k = 0$.
When this occurs, the system enters into \textit{Safety Control mode}. The safety controller (\ref{eq:emergencycontrol}) turns on, and controls the system in a safe manner until a time when a satisfactory target set is found (at which point the MPC horizon resets to $N_k = T$).
The HPL approach ensures that the system (\ref{eq:VehicleModel}) will always be in the safe set (\ref{eq:emergencyset}) when $N_k = 0$ and the safety controller (\ref{eq:emergencycontrol}) needs to turn on. We prove this in Sec.~\ref{sec:feasproof}.

\section{The HPL Algorithm}
Alg.~\ref{alg:HPL} summarizes the HPL control policy. Importantly, HPL only requires a local forecast of the new task environment, rather than the entire environment description. 
Gaussian Processes, trained offline on trajectories from past control tasks, are used to construct reduced-dimension strategy sets. 
Target sets are formed in conjunction with a safety controller, and are used as terminal sets in a shifting-horizon MPC controller.

\begin{algorithm}[h!]
\caption{HPL Control Policy}\label{alg:HPL}
\begin{spacing}{1}
\begin{algorithmic}[1]
\State \textbf{\underline{parameters:}} $d_{\mathrm{thresh}}, T, N, \mathcal{X}_E, \pi_E$
\State \textbf{\underline{input:}} $f$, $\mathcal{X}$, $\mathcal{U}$, $\{\mathrm{Ex}(\mathcal{T}^1, \bm{\theta}^1),..., \mathrm{Ex}(\mathcal{T}^n, \bm{\theta}^n)\}, ~ \bm{\theta}^{n+1}$
\State \textbf{\underline{output:}} $\mathrm{Ex}(\mathcal{T}^{n+1}, \bm{\theta}^{n+1})$
\State
\State \textbf{\underline{offline:}} 
\State \textbf{train} GPs using stored executions as in Sec.\ref{ssec:trainingGP}
\State
\State \textbf{\underline{online:}} 
\State \textbf{initialize} $k=0$, $N_k = T$, $\mathrm{SetList} = [~]$
\For{each time step $k$}
\State \textbf{collect} $(x_k, \theta^{n+1}_{k:k+N})$
\State \textbf{find} $[\tilde{\mathcal{X}}_{k+T}, \tilde{\mathcal{U}}_{k : k+T}, \mathcal{C}_k]$ using (\ref{eq:gpeq}) - (\ref{eq:strategystateconstraint})
\State  \textbf{construct} $\mathcal{X}_{k+T}$ using (\ref{eq:xt})
\If{$\mathcal{C}_k < d_{\mathrm{thresh}}$}
\State $\mathcal{X}_{k+T} = [~]$
\EndIf
\State \textbf{append} $\mathcal{X}_{k+T}$ to $\mathrm{SetList}$ (\ref{eq:setlist}) and shift sets

\If{all sets in $\mathrm{SetList}$ (\ref{eq:setlist}) are \textbf{empty}}
\State $u_k = \pi_e(x_k)$
\Else 
\State \textbf{calculate} $N_{k}$ using (\ref{eq:nk})
\State \textbf{solve} MPC (\ref{eq:lowlevelmpc}) with horizon $N_{k}$
\State $u_k = u_{0|k}^\star$ (\ref{eq:lowlevelinput})
\EndIf
\EndFor

\State \textbf{end}
\end{algorithmic}
\end{spacing}
\end{algorithm}

\section{Feasibility Proof}\label{sec:feasproof}
We prove that Alg.~\ref{alg:HPL} outputs a feasible execution for a new control task $\mathcal{T}^{n+1}$.

\begin{theorem}\label{th1}
	Let Assumption~\ref{ass:safetycontroller} hold.
	Consider the availability of feasible executions (\ref{eq:iterationvecs}) by a constrained system (\ref{eq:VehicleModel})-(\ref{eq:VehicleModelConstr}) of a series of control tasks $\{\mathcal{T}^1, \dots, \mathcal{T}^n \}$ in different environments $\{E(\bm{\theta}^1), \dots, E(\bm{\theta}^n)\}$.
	Consider a new control task $\mathcal{T}^{n+1}$ in a new environment $E(\bm{\theta}^{n+1})$. 
	If $x_0^{n+1} \in \mathcal{X}_E$, then the output of Alg.~\ref{alg:HPL} is a feasible execution of $\mathcal{T}^{n+1}$: $\mathrm{Ex}(\mathcal{T}^{n+1}, \bm{\theta}^{n+1})$.
\end{theorem}
For ease of reading, we drop the task index $(\cdot)^{n+1}$.

\begin{proof}
We  use induction to prove that for all $k\geq 0$, the iteration loop (Lines 10-23) in Alg.~\ref{alg:HPL} finds an input $u_k$ such that the resulting closed-loop trajectory satisfies system and environment constraints. 

At time $k=0$ of the new task $\mathcal{T}^{n+1}$, the target set list can contain at most one non-empty set, $\mathcal{X}_T$ (\ref{eq:xt}). 
If $\mathcal{X}_T$ is non-empty, and the resulting (\ref{eq:lowlevelmpc}) is feasible, then there exists an input sequence $[u_{0|0}, \dots, u_{T-1|0}]$ calculated by (\ref{eq:lowlevelmpc}) satisfying all state and input constraints (\ref{eq:taskconstraints}), with $x_{T|0} \in \mathcal{X}_T$.
However, if $\mathcal{X}_T$ is empty or (\ref{eq:lowlevelmpc}) is infeasible, we instead apply the safety control law $u_0 = \pi_e(x_0)$. By assumption, $x_0 \in \mathcal{X}_E$, so this input is feasible.
Thus we have shown that the iteration loop in Alg.~\ref{alg:HPL} is feasible for $k=0$.

Next, we show that the iteration loop of Alg.~\ref{alg:HPL} is recursively feasible.
Assume that at time $k>0$, the low-level policy (\ref{eq:lowlevelmpc}-\ref{eq:lowlevelinput}) is feasible with horizon $N_k$, and let ${\bf{x}}_{k:k+N_k | k}^{\star}$ and ${\bf{u}}_{k:k+N_k-1 | k}^{\star}$ be the optimal state trajectory and input sequence according to (\ref{eq:lowlevelmpc}), such that
\begin{align}
    u_k &= u_{k|k}^{\star}\label{eq:closedloopinput}\\
    x^{\star}_{k+N_k|k} &\in \mathcal{X}_{k+N_k}. \label{eq:stateintermconstraint}
\end{align}
If at time $k+1$ a non-empty target set $\mathcal{X}_{k+T+1}$ is constructed according to (\ref{eq:xt}) such that (\ref{eq:lowlevelmpc}) is feasible, then there exists a feasible input sequence $[u_{k+1|k+1}, \dots, u_{k+T|k+1}]$ satisfying all state and input constraints such that $x_{k+T+1} \in \mathcal{X}_{k+T+1}$. 

If at time $k+1$ the target set is \textit{empty}, or (\ref{eq:lowlevelmpc}) is infeasible, we must consider two cases separately:

\underline{Case 1: The MPC horizon at time step $k$ is $N_k > 1$.}
In the absence of model uncertainty, when the closed-loop input $u_k$ (\ref{eq:closedloopinput}) is applied, the system (\ref{eq:VehicleModel}) evolves such that
\begin{align}\label{eq:statevolution}
    x_{k+1} = x^{\star}_{k+1|k}.
\end{align}
According to Alg.~\ref{alg:HPL}, when the empty target set $\mathcal{X}_{k+1+T}$ is added to the target set list (\ref{eq:setlist}), the MPC horizon is shortened and the most recent non-empty target set is used again. Since $N_k > 1$, we are guaranteed at least one non-empty target set in (\ref{eq:setlist}) that may used as a feasible terminal constraint in the low-level controller (\ref{eq:lowlevelmpc}).
At time step $k+1$, the shifted input sequence ${\bf{u}}_{k+1:k+N_M-1 | k}^{\star}$ will be optimal for this shifted horizon optimal control problem (with a corresponding state trajectory ${\bf{x}}_{k+1:k+N_M | k}^{\star}$). At time step $k+1$, Alg.~\ref{alg:HPL} applies the second input calculated at the previous time step: $u_{k+1} = u^{\star}_{k+1 | k}$.

\underline{Case 2: $N_k = 1$.} In this scenario, the target set list (\ref{eq:setlist}) at time step $k+1$ is empty, resulting in $N_{k+1} = 0$ (\ref{eq:nk}).
However, combining the fact that $N_k=1$ with (\ref{eq:stateintermconstraint}) and (\ref{eq:statevolution}), we note that
\begin{align}
    x_{k+1} \in \mathcal{X}_{k+1} \subseteq \mathcal{X}_E, \nonumber
\end{align}
by construction of the target set (\ref{eq:xt}). 
This implies that at time step $k+1$, system (\ref{eq:VehicleModel}) is necessarily in the safe set (\ref{eq:emergencyset}), and so application of the safety controller will result in a feasible input, $u_{k+1} = \pi_e(x_{k+1})$.

We have shown that \textit{i)} the online iteration loop (Lines 10-23) in Alg.~\ref{alg:HPL} finds a feasible input at time step $k=0$, and \textit{ii)} if the loop finds a feasible input at time step $k$, it must also find a feasible input at time $k+1$. 
We conclude by induction that the iteration loop in Alg.~\ref{alg:HPL} finds a feasible input $u_k \forall k \in \mathbb{Z}_{0+}$ in the new task $\mathcal{T}^{n+1}$. This results in a feasible execution of $\mathcal{T}^{n+1}$.
\end{proof}

\section{Robot Path Planning Example}

We demonstrate the effectiveness of our proposed control architecture in a robotic path planning example. 

\subsection{System and Task description}
Consider a UR5e\footnote{https://www.universal-robots.com/products/ur5-robot/} robotic arm tasked with maneuvering its end-effector through a tube with varying slope.
The UR5e has high end-effector reference tracking accuracy, allowing us to use a simplified end-effector model in place of a discretized second-order model \cite{robothard3}.
At each time step $k$, the state of the system is $z_k$,
\begin{align}
    z_k = [x_k,~ \dot{x}_k,~y_k, ~\dot{y}_k], \nonumber
\end{align}
where $x_k$ and $y_k$ are the coordinates of the end-effector, and $\dot{x}_k$ and $\dot{y}_k$ their respective velocities.
The inputs to the system at time step $k$ are 
\begin{align}\label{eq:systeminputs}
    u_k = [\ddot{x}_k,~ \ddot{y}_k],
\end{align}
the accelerations of the end-effector in the $x$ and $y$ direction, respectively.
We model the base-and-end-effector system as a quadruple integrator:
\begin{align}
    {z}_{k+1} & = A z_k + B u_k \label{eq:simplifiedRobotModel}\\ 
    A &=
    \begin{bmatrix}1 & dt & 0 & 0 \\ 0 & 1 & 0 & 0\\ 0 & 0 & 1 & dt \\ 0 & 0 & 0 & 1 \end{bmatrix}, B = \begin{bmatrix}0 & 0 \\ dt & 0 \\ 0 & 0 \\ 0 & dt \end{bmatrix}, \nonumber
\end{align}
where $dt = 0.01$ seconds is the sampling time. This model holds as long as we operate within the region of high end-effector reference tracking accuracy, characterized experimentally as the following state and input constraints:
\begin{equation}\label{eq:taskinputconstr}
\begin{aligned}
    {\mathcal{X}} &=  \begin{bmatrix}-3 \\ -3\end{bmatrix} \leq  \begin{bmatrix}\dot{x}_k \\ \dot{y}_k  \end{bmatrix}  \leq   \begin{bmatrix}3  \\ 3 \\ \end{bmatrix}  \nonumber \\
    {\mathcal{U}} &=  \sqrt{\ddot{x}_k + \ddot{y}_k} \leq 1, \nonumber
\end{aligned}
\end{equation}
where the states $x_k$ and $y_k$ are not constrained by the system, but by the particular task environment.

Each control task $\mathcal{T}^i$ requires the end-effector to be controlled through a different tube, described using the environment descriptor function $\bm{\theta}^i$, as quickly as possible.
Here, the function $\bm{\theta}^i$ maps a state in a tube segment to the slope of the constant-width tube. Different control tasks $\{\mathcal{T}^1, \dots, \mathcal{T}^n \}$ correspond to maneuvering through tubes of constant width but different piecewise-constant slopes. 

We choose two strategy states for these tasks: 
\begin{align}
    \tilde{x}_k = [s_k, ~h_k], \nonumber
\end{align}
where $s_k$ is the cumulative distance along the centerline of the tube from the current point $(x_k, y_k)$ to the projection onto the centerline, and $h_k$ is the distance from $(x_k, y_k)$ to the centerline. The strategy states therefore measure the total distance traveled along the tube up to time step $k$, and the current signed distance from the center of the tube. 
The system inputs (\ref{eq:systeminputs}) are used as strategy inputs.
\begin{figure}
    \centering
    \includegraphics[width=\columnwidth]{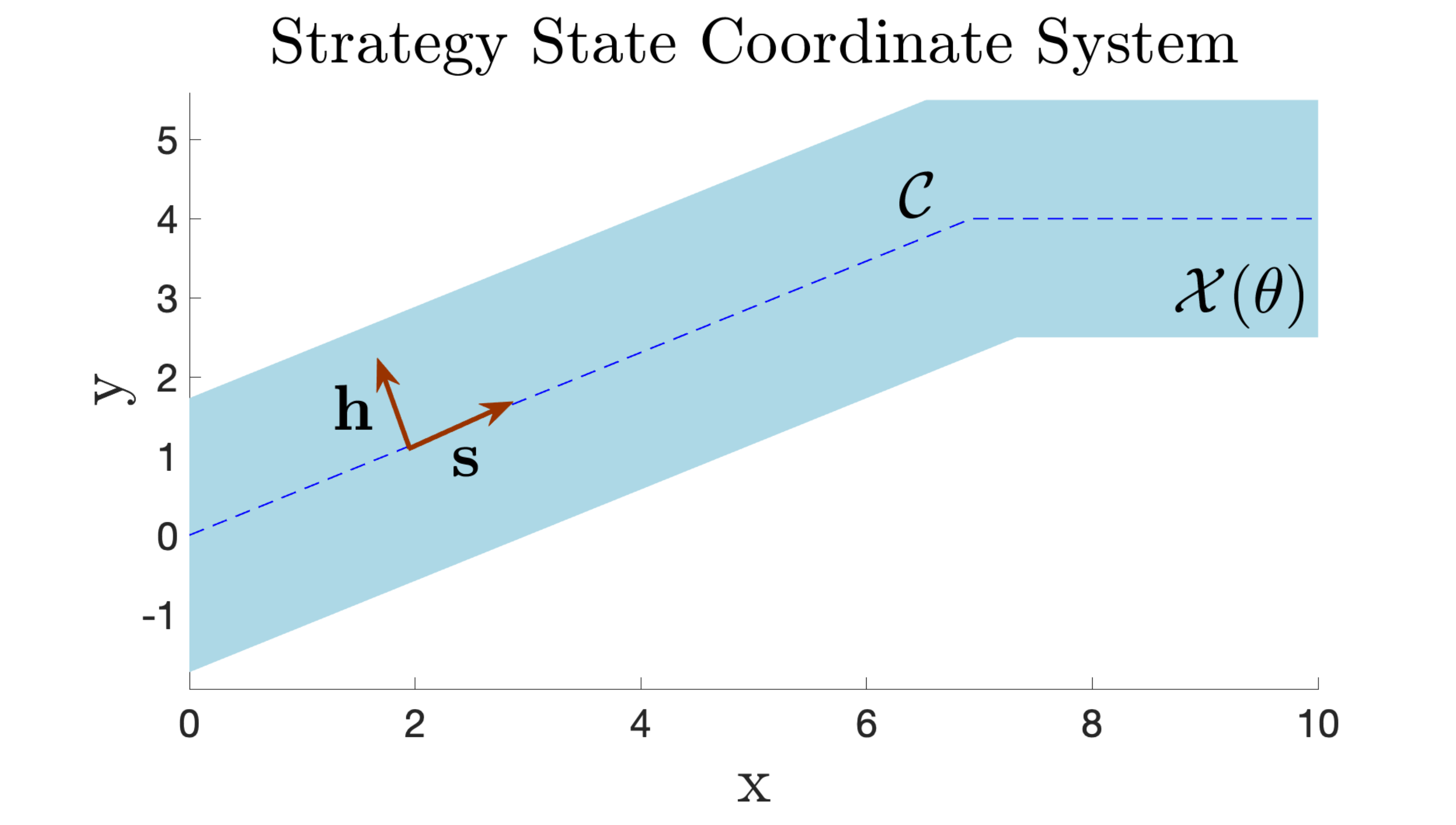}
    \caption{The end-effector is constrained to stay in the light blue tube $\mathcal{X}(\bm{\theta})$, whose centerline is plotted in dashed blue. The strategy states $s$ and $h$ measure the cumulative distance along the centerline and the distance from the centerline.}
    \label{fig:coordinates}
\end{figure}

Our safety controller (\ref{eq:emergencycontrol}) is an MPC controller which tracks the centerline of the tube at a slow, constant velocity of $0.5$ meters per second. The system (\ref{eq:simplifiedRobotModel}) in closed-loop with this centerline-tracking controller is able to solve each of the considered tasks without breaking the state constraints imposed by the environment (i.e. without hitting the tube boundary).
The safe set $\mathcal{X}_E$ (\ref{eq:emergencyset}) is determined offline using sampling-based forward reachability.

\subsection{Hierarchical Predictive Learning Results}
\begin{figure}
    \centering
    \includegraphics[width=\columnwidth]{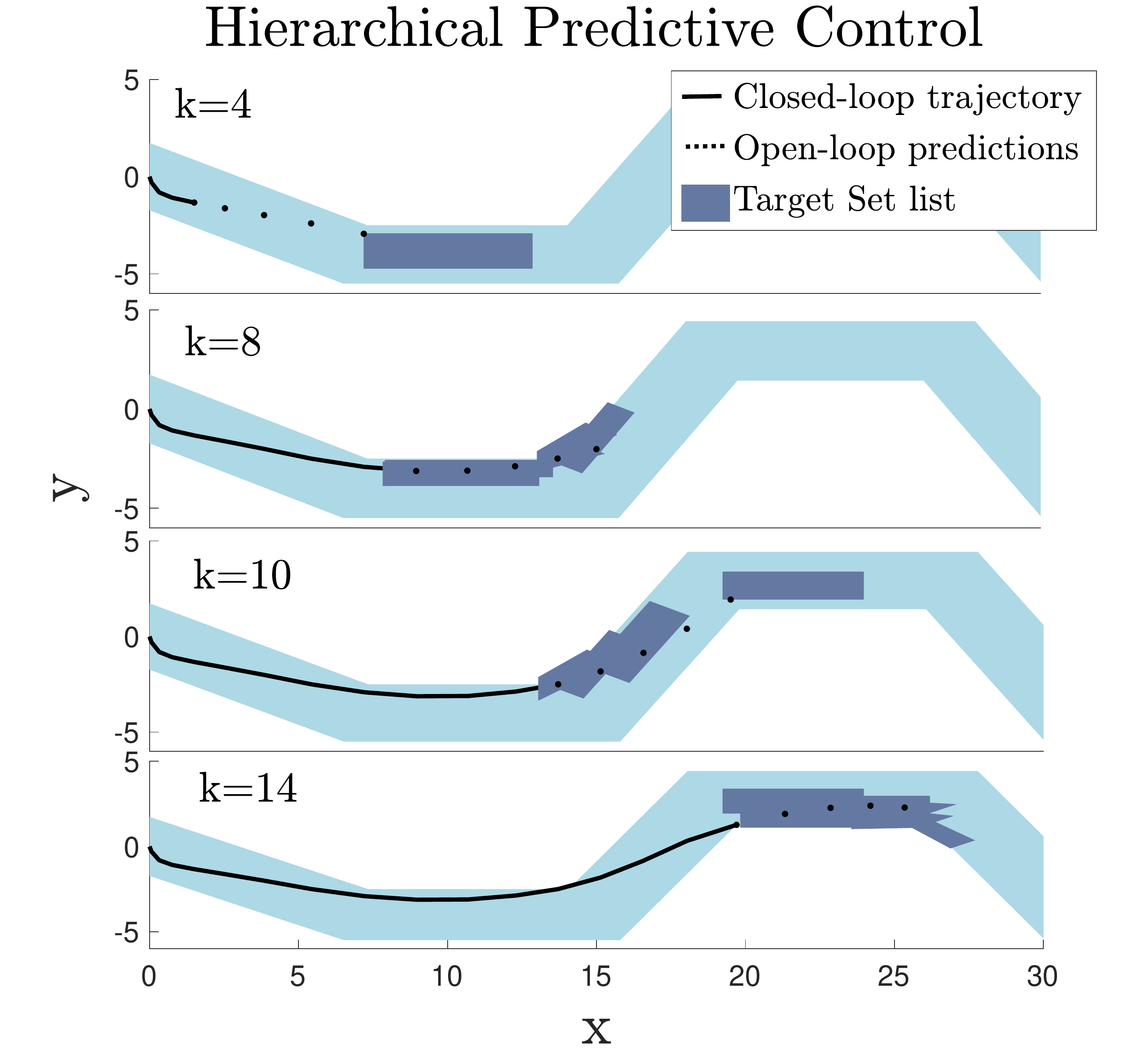}
    \caption{At each time step, the target set list (\ref{eq:setlist}) provides different regions in the task space for the system to track. 
    }
    \label{fig:plotprogression}
\end{figure}

\begin{figure}
    \centering
    \includegraphics[width=\columnwidth]{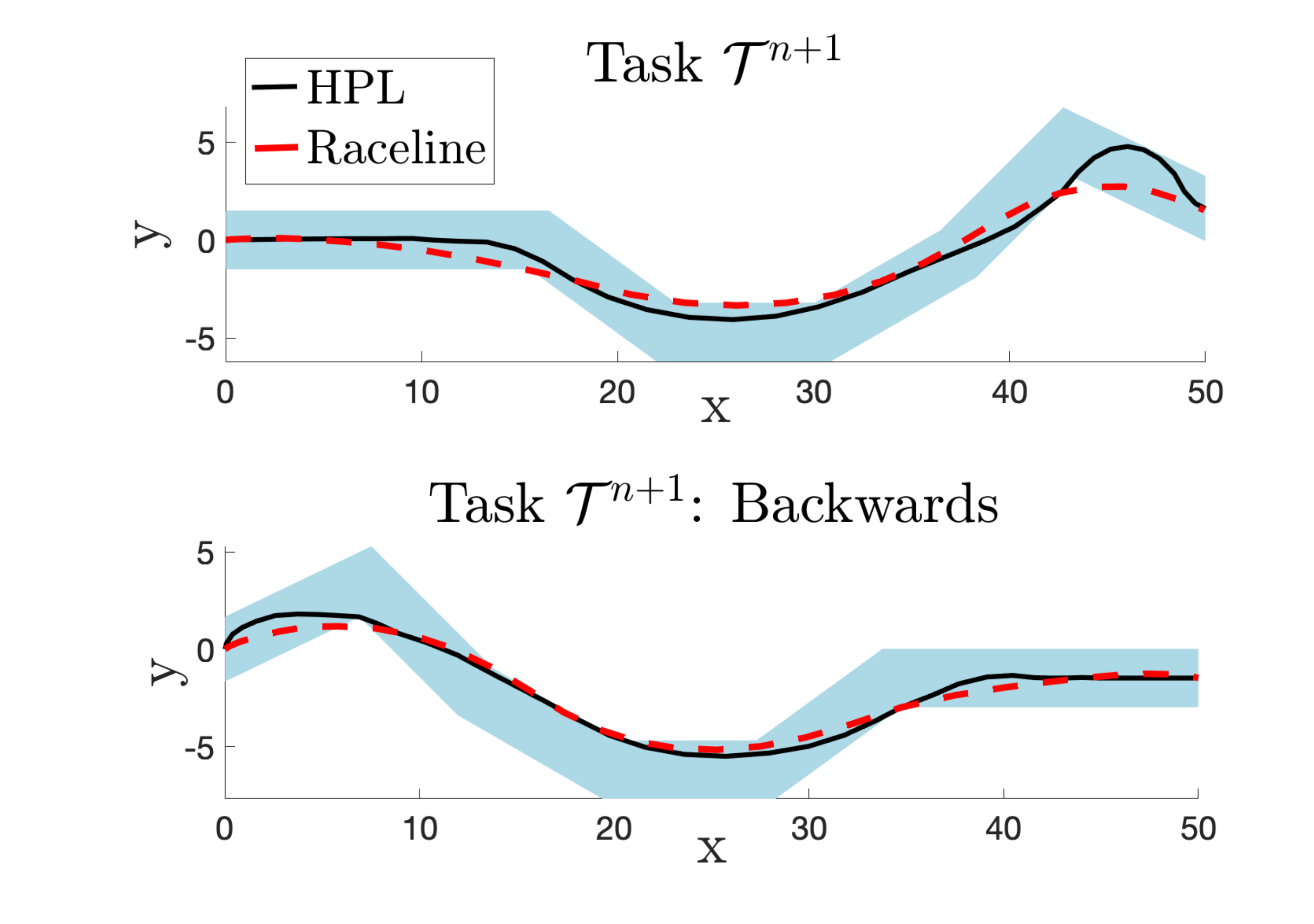}
    \caption{The HPL execution is compared to the raceline (the fastest possible execution), as determined by an LMPC \cite{rosolia2016learning}. Respective execution times in [s] are $6.5$ (LMPC), $8.8$ (HPL), and $12.8$ (Centerline-tracking $\pi_e$).}
    \label{fig:bf1}
\end{figure}
We test our proposed control architecture in simulation.
We begin by collecting executions that solve a series of $20$ control tasks $\{ \mathcal{T}^1, ...,\mathcal{T}^{20}\}$, with each control task corresponding to a different tube shape. 
The executions are used to create training data for our GPs, using an environment forecast horizon of $N=10$ and a control horizon of $T=5$.
GPs are then found to approximate the strategies learned from solving the $20$ different control tasks. 

Figure \ref{fig:plotprogression} shows the closed-loop trajectory and target set list at various time steps of solving a new task, $\mathcal{T}^{21}$, using the HPL framework. At each time step, the final predicted state lies within the last set in the target set list; the other predicted states track any non-empty target sets as closely as possible. 
The formulation allows us to visualize what strategies have been learned, by plotting at each time step where the system thinks it should go. Indeed, we see that the system has learned to maneuver along the insides of curves, and even takes the direct route between two curves going in opposite directions.

Figure \ref{fig:bf1} shows the resulting executions for a new task solved forwards and backwards. 
We emphasize that HPL generalizes the strategies learned from training data to unseen tube segments. 
Specifically, for the tasks shown here the GPs were trained on executions solving tasks in the `forward' direction, i.e. constructed left to right using tube segments as shown in the top images.
When the tasks are solved backwards, the environment descriptors are piecewise mirror images of previously seen environment descriptors. 
For example, tube segments of certain slopes had only been traversed upwards in previous control tasks, never down.
The HPL control architecture was able to handle this change very well. 
As in Fig.~\ref{fig:plotprogression}, the forward and backwards trajectories demonstrate good maneuvering strategies, including moving along the insides of curves and cutting consecutive corners.

\section{Conclusion}
A data-driven hierarchical predictive learning architecture for control in unknown environments is presented. 
The HPL algorithm uses stored executions that solve a variety of previous control tasks in order to learn a generalizable control strategy for new, unseen tasks. 
Based on a local description of the task environment, the learned control strategy proposes regions in the state space towards which to aim the system at each time step, and provides input constraints to guide the system evolution according to previous task solutions.
We prove that the resulting policy is guaranteed to be feasible for the new tasks, and evaluate the effectiveness of the proposed architecture in a simulation of a robotic path planning task. 
Our results confirm that HPL architecture is able to learn applicable strategies for efficient and safe execution of unseen tasks. 

\section{Appendix}
We use Gaussian Processes trained on data (\ref{eq:iterationvecs}) from previous tasks to construct strategy sets for solving a new task.
Specifically, GPs use a similarity measure known as ``kernel" between input vectors to learn a nonlinear approximation of the underlying input-output mapping.  
The kernel function represents the learned covariance between two function evaluations.
In this paper we use the squared-exponential kernel for our GPs. 
Given two entries of {\bf{z}} in (\ref{eq:dataset}), this kernel evaluates as
\begin{align}
    k(z^o_i, z^w_j) = \sigma_f^2 \exp{-\frac{1}{2} \sum_{m=1}^{n_x+N+1} \frac{(z^o_{i}(m) - z^w_{j}(m))^2}{\sigma_m^2}},\nonumber
\end{align}
where $z^o_{i}(m)$ is the $m$th entry of the vector $z^o_i$.
The hyperparameters $\sigma_f$, $\sigma_1$, ..., $\sigma_{n_x+N+1}$ are learned from the training data (\ref{eq:dataset}) using maximum log-likelihood regression.
We use the Matlab GP toolbox for this process.

Given a new query vector $z^{n+1}_k$, hyperrectangular strategy sets are formed using the mean $\mu$ and variance $\sigma^2$ of the resulting Gaussian distribution:
\begin{align}
    \mu(z^{n+1}_k) & = {\bf{k}}(z^{n+1}_k)\bar{K}^{-1}{\bf{y}} 
    \label{eq:meaneval}\\
    \sigma(z^{n+1}_k)^2 &= k(z^{n+1}_k,z^{n+1}_k) - {\bf{k}}(z^{n+1}_k)\bar{K}^{-1}{\bf{k}}^\top(z^{n+1}_k),\nonumber
\end{align}
where 
\begin{align}
    {\bf{k}}(z^{n+1}_k) = [k(z^{n+1}_k, z^1_0), \dots, k(z^{n+1}_k, z^n_{b_n})],\nonumber
\end{align}
and the matrix $\bar{K}$ is formed out of the covariances between training data samples such that
\begin{align}
    \bar{K}_{i,j} = k({\bf{z}}_i, {\bf{z}}_j). \nonumber
\end{align}
The strategy sets are then constructed according to (\ref{eq:strategystateconstraint}).

\renewcommand{\baselinestretch}{0.9}
\bibliographystyle{IEEEtran}
\bibliography{IEEEabrv,main}

\end{document}